# Overview of the DHCAL Project

José Repond[1]

1 – Argonne National Laboratory
9700 South Cass Avenue, Argonne, IL 60439 – U.S.A.

On behalf of the DHCAL Collaboration

We review the status and plans of the Digital Hadron Calorimeter with Resistive Plate Chambers project.

## 1  Overview and current status

The aim of the DHCAL project is to develop a finely segmented hadron calorimeter optimized for the application of Particle Flow Algorithms (PFAs) [1] to the measurement of hadronic jets at a future lepton collider. The active element of the calorimeter was chosen to be thin Resistive Plate Chambers (RPCs) with glass as resistive plates. The readout consists of 1×1 cm$^2$ pads located on the anode side of the chambers. The electronic readout is optimized for the readout of large number of channels and is simplified to record only binary (or digital) information for each channel.

The 1$^{st}$ stage of the project consisted of developing RPCs optimized for use in a large hadron calorimeter. This stage was completed with successful tests of RPCs based on three different designs. The results [2] were published in a refereed journal.

The 2$^{nd}$ stage of the project consisted of a small prototype calorimeter which included the entire digital readout chain and was thus named Vertical Slice Test. The prototype calorimeter was subject of extensive testing with both cosmic rays and in particle beams. The results were summarized in five papers [3-7], published in a refereed journal. In addition, the first PhD thesis [8] of the project was written on the environmental studies of the RPC performance.

The 3$^{rd}$ stage of the project consists of the preparation for and construction of a large scale prototype calorimeter with approximately 400,000 readout channels. In preparation for the construction, extensive R&D has been carried out, which was completed about a year ago. The construction of the prototype calorimeter is ongoing and about 2/3 done [9].

The 4$^{th}$ stage of the project consists of preparation for a technical prototype calorimeter, including all the bells and whistles necessary for a colliding beam detector. This stage has not begun in earnest yet. Nevertheless, we shall present some initial thoughts about the challenges ahead.

The project is being carried out by the DHCAL collaboration [10] counting 36 heads from Argonne National Laboratory, Boston University, Fermilab, University of Iowa and University of Texas at Arlington. This research is being performed under the auspices of the CALICE collaboration [11], which develops imaging calorimetry for the application of PFAs to the measurement of hadronic jets at a future lepton collider.

## 2  The prototype calorimeter

Construction of the prototype calorimeter is in full swing [9]. Table 1 summarizes the current status. Needless to say, that the construction process has been a great learning experience. The RPCs are built with tolerances at the 100 μm level. Despite the use of high precision assembly fixtures, the RPC assembly turned out to be significantly more labor intensive than expected. Therefore, the current assembly procedure is not viable for the construction of a large calorimeter.

For the construction of the electronic readout system we opted to follow a conservative approach, with extensive testing at every step. This approach is time consuming, but necessary when assembling a large system, where malfunctions at the percent level might compromise the reliable operation of the entire system. When envisaging the construction of an entire hadron calorimeter for a future lepton collider, most of the time consuming testing procedures will not be necessary, as the design will already have been finalized.

| Task | Status | Comment |
|---|---|---|
| RPC construction | 70% done | Very tedious |
| Cassette construction | Design completed<br>1st prototype assembled and tested<br>Material for more cassettes in hand, but design not yet finalized | Costly, but not very labor intensive |
| Front-end electronics | All DCAL chips tested<br>Front-end boards prototyped and fully debugged. Board fabrication completed, assembly to begin shortly | Pursued a very conservative approach |
| Back-end electronics | Data collectors 100% done<br>Timing and trigger modules redesigned, fabricated and being assembled | No or only minor changes compared to the Vertical Slice Test |
| Low Voltage supplies | Power supplies in hand<br>1st distribution box assembled and tested<br>Parts for all units in hand | |
| High Voltage supplies | Units in hand<br>Computer controlled program developed and commissioned | |
| Gas system | Gas mixing unit completed and tested<br>2nd distribution rack being assembled | |
| Cables | Required length measured<br>Production not yet started | |
| DAQ software | Implemented into CALICE framework<br>99% complete | CALICE framework needed for combined runs with other calorimeter prototypes |
| Event building | First version of event builder written | |

| Event display | Complete | |
|---|---|---|
| Data analysis | Started to reconstruct tracks in cosmic ray data<br>Calibration procedure still to be developed | Large experience from Vertical Slice Test will be useful |
| Simulation | RPC response simulated<br>Implementation of DHCAL geometry into MOKKA framework ongoing | |

Table 1: Summary of the status of the prototype construction effort as of April 2010.

Table 2 summarizes the plans and time scales concerning the construction and testing of the prototype calorimeter. The combined tests with the ECAL allude to tests with the CALICE Silicon-Tungsten electromagnetic calorimeter prototype [12].

| Task | Dates | Comments |
|---|---|---|
| Construction | Complete by June 30$^{th}$ | Should not slip much more… |
| Cosmic Ray testing of cubic meter | April through August | |
| Installation into MTest | September | |
| 1$^{st}$ data taking period | October | DHCAL standalone (with TCMT) |
| 2$^{nd}$ data taking period | December | Combined tests with ECAL |
| 3$^{rd}$ data taking period | Early 2011 | DHCAL standalone or combined |
| Disassembly and shipping of stage | March 2011 | (Maybe not so) hard deadline |

Table 2: Plans and time scales for the prototype calorimeter construction and testing.

## 3  R&D beyond the prototype calorimeter

Further R&D is needed to develop a viable design of a DHCAL with RPC calorimeter module, which could be inserted into a colliding beam detector. Following, we list a few key areas for future work:

a) 1-glass RPCs: Argonne developed a new RPC design which is based on a single glass plate instead of the traditional two plates. Prototype chambers utilized in the Vertical Slice Test performed very well. These chambers offer various distinct advantages compared to the standard 2-glass design: constant pad multiplicity around unity, smaller overall thickness, better rate capability etc. In order to pursue this design further, we will assemble a few 1-glass

RPCs with the electronics of the prototype calorimeter and test them with cosmic rays and possibly in a particle beam.

b) The next version of the DCAL [13] front-end chip requires a major redesign. Rather than explore the potential of power pulsing (with its obvious drawbacks for use in test beams and with cosmic rays), we plan on pursuing recent developments in ultra-low power circuitry. Other possible improvements compared to the current version of the front-end chip are an increased channel count, token ring passing and higher readout speeds (mostly for other than the present applications).

c) Currently the front-end board and pad-boards are separate entities connected by drops of conductive glue. This design avoids the use of costly blind or buried vias, but comes at the expense of a thicker readout layer and extra effort to mate the two boards. A new solution needs to be developed which consists of a single board without buried or blind vias, but which nevertheless provides the required low noise characteristics.

d) Both low voltage and high voltage will need to be distributed to the individual layers and readout boards of a module. Ways to generate, control, and monitor the power for each layer within a module need to be developed. Particularly promising might be the application of Cockcroft-Walton technologies.

e) Gas distribution within a module is a challenging task. Ways to ensure uniform flow in all layers need to be developed. But even more challenging is the need to recycle the gas, which we will address in the next chapter.

## 4  Gas recycling

Our RPCs are operated in avalanche mode with a mixture of three gases. This mixture has been optimized by the RPC community over the past two decades or so. Table 3 lists the gases together with their Global Warming Potential (GWP). As is evident from the table, these gases are damaging the environment.

| Gas | Fraction [%] | Global warming potential (100 years, $CO_2$=1) | Fraction*GWP |
| --- | --- | --- | --- |
| Freon R134a | 94.5 | 1430 | 1351 |
| Isobutan | 5.0 | 3 | 0.15 |
| $SF_6$ | 0.5 | 22,800 | 114 |

Table 3: Gas mixture for operating RPCs in avalanche mode.

The 40 layer prototype calorimeter features a gas volume of about 40 liters. In the testbeam the higher rate of avalanches necessitates at least 10 volume changes per day [7]. Operating for an approximate four months requires therefore 48,000 liters of mixed gas. This gas will be released into the atmosphere. A simple calculation shows that the corresponding greenhouse gas emission corresponds to driving 30 cars around the globe. This is not good, but also not disastrous for the environment.

Extrapolating to the hadron calorimeter of a future lepton collider introduces a factor of about 100 in gas volume and, say, a factor of 20 due to the extended data taking period. In consequence, the emission would correspond to driving about 50,000 cars around the globe.

Apart from the high material cost related to the one-time use of these (expensive) gases, the disastrous effect on the planet is clearly not acceptable.

Two methods of recycling RPC gases are being developed. The closed loop option pursued at CERN utilizes filters to eliminate the contaminants from the gas [14]. The open loop option pursued by the INO experiment utilizes cold traps set to different temperatures to condense the gases one-by-one [15]. The thus recovered gases are then remixed to provide the standard RPC mixture.

Both methods are still in the development stage. We have established initial contact with the INO experiment with the prospect of future collaboration on this issue.

## 5  Acknowledgements

The author would like to thank the organizers for an impeccably organized workshop and for the opportunity to present the work of the DHCAL collaboration to the International Linear Collider community.